\documentstyle[amssymb,aps]{revtex}

\begin{document}
\author{Qing-yu Cai}
\title{Optimal Experimental Scheme for Bennett-Brassard 1984 Quantum Key
Distribution Protocol with Weak Coherent Sources, Noisy and Lossy Channel }
\address{State Key Laboratory of Magnetic Resonances and Atomic and Molecular\\
Physics, Wuhan Institute of Physics and Mathematics, the Chinese Academy of\\
Sciences, Wuhan 430071, China}
\maketitle

\begin{abstract}
It is the first scheme which allows the detection apparatus to achieve both
the photon number of arriving signals and quantum bit error rate of the
multiphoton pulses precisely. We show that the upper bound of the fraction
of the tagged multiphoton pulses counts is $\mu $, which is independent of
the channel loss and the intensity of the decoy source. Such upper bound is $%
inherent$ and cannot be reduced any longer as long as the weak coherent
scouces and high lossy channel are used. We show that our scheme can be
implemented even if the channel loss is very high. A stronger intensity of
the pulse source is allowable to improve the rate of quantum key
distribution. Both the signal pulses and decoy pulses can be used to
generate the raw key after verified the security of the communication. We
analyze that our scheme is optimal under today's technology.
\end{abstract}


\smallskip Quantum key distribution (QKD) is a physically secure method, by
which private key can be created between two partners, Alice and Bob, who
share a quantum channel and a public authenticated channel [1]. The key bits
then be used to implement a classical private key cryptosystem, or more
precisely called $one-time$ $pad$ algorithm, to enable the partners to
communicate securely.

An optical QKD system includes the photon sources, quantum channels,
single-photon detectors, and quantum random-number generators. In principle,
optical quantum cryptography is based on the use of single-photon Fock
states. However, perfect single-photon sources are difficult to realize
experimentally. Practical implementations rely on faint laser pulses in
which photon number distribution obeys Possionian statistics. So, Eve can
get benefits from the multiphoton pulses. If the quantum channel is high
lossy, Eve can obtain full information of the final key by using photon
number splitting (PNS) attack without being detected. In GLLP [2], it has
been shown that the secure final key of BB84 protocol [3] can be extracted
from sifted key at the asymptotic rate 
\begin{equation}
R=(1-\Delta )-H_{2}(e)-(1-\Delta )H_{2}(\frac{e}{1-\Delta }),
\end{equation}
where $e$ is the quantum bit error rate (QBER) found in the verification
test and $\Delta =p_{M}/p_{D}$, where $p_{M}$ is the probability of
detecting a multiphoton pulse and $p_{D}$ is the probability that an emitted
photon is detected. This means that both the QBER $e$ and the fraction of
tagged signals $\Delta $ are important to generate the secure final key. It
has been shown that Eve's PNS attack will be limited when Alice and Bob use
the decoy-state protocols [4-6]. In such decoy-state protocols, the
detection apparatus cannot resolve the photon number of arriving signals.
Thus, Eve may use a photon number splitting and resending (PNSR) attack on
the multiphoton pulses, i.e., Eve replaces one photon of the multiphoton
pulses by a false one and forwards the pulses to Bob, to eavesdrop Alice's
information. Therefore, it requires that the communication partners should
have the capacity to achieve the QBERs of multiphoton pulses. As a matter of
fact, Eve's some other attacks, such as coherent multiphoton pulse attack,
should also be considered or else security of the final key will be
unreliable.

In this paper, we present an experimental detection apparatus which allows
Bob to achieve the photon number of arriving signals and QBER of the
multiphoton pulses precisely. The upper bound of the information Eve can
gain from channel loss is $\mu $, no matter how high the channel loss is. We
show that all the multiphoton pulses verified by Bob should be discarded in
the end. As a consequence, security of the QKD is only determined by the
QBER and the intensity of the pulse sources $\mu $, irrespective of the
channel loss efficiency and the intensity of the decoy sources. Finally, we
discuss and conclude that our scheme is optimal under today's technologies.

At present, practical ``single-photon'' sources rely on faint laser pulses
in which photon number distribution obeys Possionian statistics. Most often,
Alice sends to Bob a weak laser pulse in which she has encoded her bit. Each
pulse is a priori in a coherent state $|\sqrt{\mu }e^{i\theta }\rangle $ of
weak intensity. Since Eve and Bob have no information on $\theta $, the
state reduces to a mixed state $\rho =\int \frac{d\theta }{2\pi }|\sqrt{\mu }%
e^{i\theta }\rangle \langle \sqrt{\mu }e^{i\theta }|$ outside Alice's
laboratory. This state is equivalent to the mixture of Fock state $%
\sum_{n}p_{n}|n\rangle \langle n|$, with the number $n$ of photons
distributed as Possionian statistics $p_{n}=P^{\mu }(n)=\mu ^{n}e^{-\mu }/n!$%
. The source that emits pulses in coherent states $|\sqrt{\mu }e^{i\theta
}\rangle $ is equivalent to the representation as below: With probability $%
p_{0}$, Alice does nothing; With probability $p_{n}$ $(n>0)$, Alice encodes
her bit in $n$ photons. Thus, Eve can use two different eavesdropping
strategies to gain information about Alice's qubit. Eve first performs a
nondemolition measurement to gain the photon number of the laser pulses.
When she finds there is only one photon in the pulses, she implements
symmetric individual (SI) attacks to gain Alice's information [7].
Otherwise, if there are two or more than two photons in the pulses, she will
perform some multiphoton attacks on Alice's qubit, e.g., she may implement
PNS attack on Alice's qubit. The probability of that a nonempty pulse
contains more than one photon is that $p(n>1|n>0,\mu )=\frac{1-p(0,\mu
)-p(1,\mu )}{1-p(0,\mu )}=\frac{1-e^{-\mu }(1+\mu )}{1-e^{-\mu }}%
\thickapprox \frac{\mu }{2}$. It is a fact that the probability of $%
p(n>1|n>0,\mu )$ can be made arbitrary small so that weak pulses are
practical and have indeed been used in the vast majority of experiments [1].
However, in long distance QKD, the channel transmittance $\eta $ [8] can be
rather small. If $\eta <(1-e^{-\mu }-\mu e^{-\mu })/\mu $, Eve can gain full
information of Bob's final key by using the PNS attack [9].

$Photon$ $number$ $measurement$ $and$ $QBERs$ $of$ $multiphoton$ $pulses$.
---In Fig. 1, we present a method to obtain the photon number of the
multiphoton pulses. Furthermore, QBERs of different multiphoton pulses can
also be achieved. In Fig. 1(A), there are $N$ beam splitters (BSs) at the
end of the fiber. The probability that two photons were detected in one
single-photon detector is approximate to $O(1/N)$ [10]. Bob can perfectly
achieve the photon number of the multiphoton pulses by increasing the number
of the BS. In Fig. 1(B), there is a polarization beam splitter (PBS) at each
output of each BS. Obviously, these PBSs should be parallel. Using Alice's
public announcement, Bob can obtain the QBER with his statistic data. In an
uncharacteristic channel, all QBERs of different multiphoton pulses should
be identical. Thus, if either the photon number statistics results or the
QBER is abnormal, Alice and Bob will know that Eve is in line.

$Eavesdropping$ $hidden$ $in$ $channel$ $losses$. ---In order to detect
Eve's PNS attack, Alice and Bob will use the $improved$ decoy-state protocol
to verify the security of their communication. Suppose Alice and Bob select $%
\mu $ as signal source and $\mu ^{\prime }$ as the decoy source in our
scheme. Without Eve's presence, photon number distributions are also
Poissonian with the channel transmittance $\eta $, 
\begin{eqnarray}
P_{loss}^{\mu }(n) &=&\frac{(\eta \mu )^{n}}{n!}e^{(-\eta \mu )}, \\
P_{loss}^{\mu ^{\prime }}(n) &=&\frac{(\eta \mu ^{\prime })^{n}}{n!}%
e^{(-\eta \mu ^{\prime })}.
\end{eqnarray}
In general, let us assume that Eve has a lossless quantum channel. She
implements PNS attack on the photon pulses with probability $P_{Eve}(n)$.
Essentially, the idea of decoy-state is that [6] 
\begin{eqnarray}
P_{Eve}(signal) &=&P_{Eve}(decoy)=P_{Eve}(n) \\
e_{n}(signal) &=&e_{n}(decoy)=e_{n}.
\end{eqnarray}
The sufficient condition that Eve can redistribute the photon number without
being detected is that 
\begin{eqnarray}
P^{\mu }(n)[1-P_{Eve}(n)]+P^{\mu }(n+1)P_{Eve}(n+1) &=&P_{loss}^{\mu }(n), \\
P^{\mu ^{\prime }}(n)[1-P_{Eve}(n)]+P^{\mu ^{\prime }}(n+1)P_{Eve}(n+1)
&=&P_{loss}^{\mu ^{\prime }}(n)
\end{eqnarray}
are satisfied for all $n$ with $P_{Eve}(0)=0$. $P_{Eve}(1)$ implies that Eve
blocks some single photon pulses to satisfy $P_{loss}^{\mu }(0)$. Clearly, $%
P_{Eve}(n)$ should be independent of $\mu $ because of the decoy source $\mu
^{\prime }$, vice versa. Solution of $P_{Eve}(1)$ is that 
\begin{equation}
P_{Eve}(1)=\frac{e^{-\eta \mu }-e^{-\mu }}{\mu e^{-\mu }}=1-\eta ,
\end{equation}
where we assumed that $e^{\mu (1-\eta )}=1+\mu (1-\eta )+O(\mu ^{2})$
because $\mu $ is small. This implies that the distribution of photon number
in Bob's detection apparatus can be satisfied when Eve blocks the single
photon pulses with the probability $1-\eta $, i.e., transmittance efficiency
of single photon pulses is $\eta $ on this occasion. From the equation 
\begin{equation}
\mu e^{-\mu }[1-P_{Eve}(1)]+P_{Eve}(2)\frac{\mu ^{2}}{2}e^{-\mu }=\eta \mu
e^{-\eta \mu },
\end{equation}
we can obtain that 
\begin{equation}
P_{Eve}(2)=2(1-\eta )\eta ,
\end{equation}
which is also independent of $\mu $. That is, if Eve can perform the PNS
attack with the probability $2(1-\eta )\eta $ on the two-photon multiphoton
pulses to satisfy the single-photon distribution on Bob's detection
apparatus. However, one can obtain that 
\begin{equation}
\frac{\mu ^{2}}{2}e^{-\mu }[1-P_{Eve}(2)]>\frac{(\eta \mu )^{2}}{2}e^{-\eta
\mu }.
\end{equation}
That is, even all the three-photon multiphoton pulses are blocked, the
probability that Bob detects two-photon multiphoton pulses will be
abnormally higher. Eve has to block both of the two photons of two-photon
multiphoton pulses with the probability $(1-\eta )^{2}$. Thus, the
probability that both of Alice's two photons can be detected in Bob's
detection apparatus is $\eta ^{2}$. Eve's such attacks are equivalent to
that Eve attacks every photon with the probability $1-\eta $ no matter which
pulses it belongs to. In this way, we can obtain all $P_{Eve}(n)$ $uniquely$
[11]. Moreover, such $P_{Eve}(n)$ are independent of the intensity of the
pulses sources. That is, decoy-state protocols are inefficient on this
occasion. In this case, such attacks are inevitable as long as the channel
is lossy. It is the $inherent$ property of the QKD with weak coherent source
and lossy channel.

From the discussion above, we know that the fraction of the two-photon
pulses that can be used by Eve is that $P_{Eve}(2)=2(\eta -\eta ^{2})$.
Since most of the multiphoton pulses are two-photon pulses, we can obtain
that 
\begin{equation}
\Delta =\frac{\frac{\mu }{2}2(\eta -\eta ^{2})}{\eta }\leq \mu \text{. }
\end{equation}
In practical long distance QKD, $\eta $ is very small. Thus, the upper bound
of the fraction of the multiphoton pulses that Eve can gain information
without being detected is $\mu $, which is independent of the channel loss
and the intensity of the decoy sources. Consequently, our scheme can be
applied no matter how high the channel loss is. If Alice and Bob verified
the absence of Eve, i.e., both the QBER and photon number distributions are
normal, they can use both the decoy pulses and signal pulses to generate
their raw key.

Another question is that Eve may use PNSR attack on the multiphoton pulses.
Without doubt, Eve's PNSR attack would cause some additional bit error rate
of the multiphoton pulses. Fortunately, our scheme is sensitive to the bit
error rate of multiphoton pulses $e_{n}$ [12]. Therefore, our scheme is
sensitive to Eve's PNSR attack.

$Optimal$ $eavesdropping$ $scheme$. ---In a general way, we will assume that
Eve's ability is only limited by the principles of quantum mechanics. It is
allowable that Eve holds a lossless channel and a perfect quantum memory.
Therefore, Eve can get benefits from noises and channel loss. In order to
avoid being detected, Eve will hide all of her attacks in the QBER or in the
channel loss. On the other hand, with the same QBER, Eve can get more
benefits from the multiphoton pulses than that from the single photon pulses
[13]. If Alice and Bob can discard some of the multiphoton pulses, security
of the communication will be enhanced. Bob can find out which pulses are
multiphoton pulses in our scheme. He labels such multiphoton pulses and
discards them. The probability that Bob detects a multiphoton pulse is $%
\epsilon \backsimeq \frac{\mu }{2}\eta ^{2}/$ $p_{D}\thickapprox \mu \eta /2$%
. Therefore, the optimal eavesdropping scheme Eve may use can be described
as follow. Eve implements SI attack on the single photon pulses. She
captures every photon with probability $1-\eta $ in her perfect channel. In
this case, the maximal information Eve can gain is given by 
\begin{equation}
I^{AE}=p_{SI}H_{SI}+p_{tagged}.
\end{equation}
The probability of the fraction of tagged multiphoton is that $%
p_{tagged}=\mu $. The probability that Eve uses a SI attack is that $%
p_{SI}=1-p_{tagged}$ [14]. In this way, Eve can attack on the communication
optimally without being detected.

In one way communication, the final key is secure if and only if $%
I^{AB}>I^{AE}$. In fact, $I^{AB}$ is determined by the unique variable QBER
since $I^{AB}=1-h\left( e\right) $, where $h(x)=-x\log _{2}x-(1-x)\log
_{2}(1-x)$. The maximal information Eve can gain by using SI attack is given
by $H_{SI}=1-h\left( \frac{1+2\sqrt{e-e^{2}}}{2}\right) $ [15]. In general,
security of the final key is determined by the parameter $\mu $, $\mu
^{\prime }$, and $e$. We give a numerical solution of the security with
different $\mu $ and variational $e$ in Fig. 2.

$Discussion$ $and$ $conclusion$. ---Today, quantum key distribution over 150
km of commercial Telcom fibers has been successfully performed. The crucial
issue in QKD is its security. Experimentally, the source is imperfect and
the channel is lossy and noisy. In our experimental scheme, both photon
number of arriving signals and QBER can be achieved precisely. We showed
that the upper bound of information Eve can gain from in the channel loss is
independent of $\eta $ and $\mu ^{\prime }$ [17], so that our scheme can be
used even if the channel loss is very high. In our scheme, information Eve
can gain is only determined by QBER and $\mu $. As a matter of fact, Eve's
such information can not be reduced any longer as long as the weak coherent
sources and imperfect quantum channel are used in long distance QKD, so that
our scheme is optimal under today's technology.

In summary, we have discussed the security of practical BB84 QKD protocol
with weak coherent sources, noises and high losses. We have presented a
detection apparatus to resolve both the photon number of arriving signals
and QBER to beat Eve's whatever PNS attacks. Our scheme is efficient even if
the channel loss is very high. Both the signal pulses and decoy pulses can
be implemented to generate the raw key after verified the security of the
QKD. A bigger $\mu $ is allowable to improve the rate of generating the raw
key. We have discussed that our scheme is optimal under today's technology.

$NOTE$.---After finished our work, we find out the paper [18] which
presented an experimental photon number resolving scheme using
time-multiplex technique. However, our scheme not only can be used to
resolve the photon number of multiphoton pulses, but also can be used to
achieve the QBER of multiphoton pulses.

The author wishes to thank Yong-gang Tan and Qiang Liu for their useful
discussions and comments. Yong Li should be thanked specially. This work is
supported by National Nature Science Foundation of China under Grant No.
10447140.

\section{references}

[1] N. Gisin, G. Ribordy, W. Tittel, and H. Zbinden, Rev. Mod. Phys. 74,
145-195 (2002).

[2] D. Gottesman, H.-K. Lo, N. L\"{u}tkenhaus, and J. Preskill, Quant. Inf.
Comp., 4(5), 325-360 (2004).

[3] C. H. Bennett, and G. Brassard, in Proceedings of the IEEE International
Conference on Computers, Systems and Signal Processing, Bangalore, India,
(IEEE, New York, 1984), pp.175-179.\smallskip

[4]W.-Y. Huang, Phys. Rev. Lett. 91, 057901 (2003).

[5] X.-B. Wang, Phys. Rev. Lett. 94, 230503 (2005).

[6] H.-K. Lo, X. Ma, K. Chen, Phys. Rev. Lett. 94, 230504 (2005); X. Ma, B.
Qi, Y. Zhao, H.-K. Lo, arXiv: quant-ph/0503005 (PRA, in press).

[7] It has been proved that SI attack is optimal when Eve attacks on the
single photon: N. L\"{u}tkenhaus, Phys. Rev. A 61, 052304 (2004).

[8] In this paper, the quantity $1-\eta $ includes imperfection of Bob's
single photon detectors. We assume that Eve maybe has the ability to control
the detection efficiency of Bob's apparatus, e.g., Eve may change the
wavelength of Alice's photon to be more sensitive to Bob's detectors.

[9] N. L\"{u}tkenhaus, M. Jahma, New J. Phys. 4, 44.1-44.9 (2002).\smallskip

[10] From $p_{n}=\mu ^{n}e^{-\mu }/n!$, we know that the probability $p_{n}$
decreases exponentially with the number $n$ increasing. So the probability
that one signal pulse contains more than two photons is very small, i.e.,
most multiphoton pulses only contain two photons. In this case, the photon
number resolving power is $O(1/N)$.

[11] In fact, we can get all $P_{Eve}(n)$ uniquely as follow. The necessary
and sufficient condition that Eve can redistribution the photon number
without being detected is that both $P^{\mu
}(n)[1-\sum_{i=0}^{n-1}f(i)]+\sum_{j=n+1}^{\infty }P^{\mu
}(j)f(n)=P_{loss}^{\mu }(n)$ and $P^{\mu ^{\prime
}}(n)[1-\sum_{i=0}^{n-1}f(i)]+\sum_{j=n+1}^{\infty }P^{\mu ^{\prime
}}(j)f(n)=P_{loss}^{\mu ^{\prime }}(n)$ are satisfied for all $n$, where $%
f(i)$ is the probability that Eve forwards $i$ photon(s) to Bob and captures 
$n-i$ photon(s) when the photon number is $n$. Using the Taylor series, we
can get that $P_{loss}^{\mu }(n)=\frac{(\eta \mu )^{n}}{n!}%
\sum_{i=0}^{\infty }\frac{(-\eta \mu )^{i}}{i!}$ and $P^{\mu }(n)=\frac{(\mu
)^{n}}{n!}\sum_{i=0}^{\infty }\frac{(-\mu )^{i}}{i!}$. When we apply such
Taylor series to these equations, we can get every $f(i)$ uniquely. These $%
f(i)$ just correspond to the case that Eve captures every photon with the
probability $1-\eta $, without reference to the value of $\mu $ and $\mu
^{\prime }$. In principle, Eve's attacks discussed above can be realized
physically. Eve only needs a lossless channel and a BS with the reflection
efficiency $1-\eta $ and the transmission efficiency $\eta $ to split
Alice's photon. A detailed discussion will be presented soon.

[12] The $e_{n}$ comes from two parts, erroneous detections and background
contribution. The background rate $e_{dark}$ is 10$^{-5}$ typically. The
probability that an arriving signal including two photons in Bob's detection
apparatus is about $\eta \mu /2$ ( typically 10$^{-4}$). In this case,
precision of $e_{n}$ will increase when the detector number $N$ in our
detection apparatus increases. Certainly, the background rate will increase
when the number of detector increases. This requires that $N\cdot e_{dark}$ $%
<\mu \eta /2$.

[13] On this occasion, we think the most general attack is the coherent
multiphoton (CMP) attack. Eve's SI attack can be described by a unitary
transformation with ancilla: U$|0\rangle _{A}|0\rangle _{E}=|0\rangle
_{A}|0\rangle _{E}$, U$|1\rangle _{A}|0\rangle _{E}=$ $\alpha |1\rangle
_{A}|0\rangle _{E}+\beta |0\rangle _{A}|1\rangle _{E}$. Both the information
Eve can gain and QBER are determined by $\alpha $ ($|\alpha |^{2}+|\beta
|^{2}=1$). Eve's CMP attack can be described as: U$\otimes $U$(|0\rangle
_{A}|0\rangle _{E}|0\rangle _{A}|0\rangle _{E})=$ $(|0\rangle _{A}|0\rangle
_{A})(|0\rangle _{E}|0\rangle _{E})$, U$\otimes $U$(|1\rangle _{A}|0\rangle
_{E}|1\rangle _{A}|0\rangle _{E})=$ $(\alpha |1\rangle _{A}|0\rangle
_{E}+\beta |0\rangle _{A}|1\rangle _{E})$ $(\alpha |1\rangle _{A}|0\rangle
_{E}+\beta |0\rangle _{A}|1\rangle _{E})$. Then Eve performs her measurement
on ancilla. If her measurement outcome is $|0\rangle |0\rangle $, she
guesses that state of Alice's qubit is $|0\rangle $. Or else, if she finds
that her measurement outcome is $|0\rangle |1\rangle $, $|1\rangle |0\rangle 
$, or $|1\rangle |1\rangle $, she knows that Alice prepares the qubit in
state $|1\rangle $. Clearly, information Eve can gain is determined by the
parameter $\alpha ^{2}$ because that $(|0\rangle _{E}|0\rangle _{E})$ are
orthogonal to the other states of ancilla. QBER in these attacks are
identical because of the same U. Thus, it seems that Eve's CMP attack is
equivalent to that she implements SI attack on each photon in the pulse
respectively. A detailed calculation will be presented soon. Some
correlative works can be found in: M. Curty and L\"{u}tkenhaus, Phys. Rev. A
69, 042321 (2004); Armand Niederberger, Valerio Scarani, Nicolas Gisin,
Phys. Rev. A 71, 042316 (2005).

[14] Transmission of two-photon pulses can be described as follow: Both two
photon were blocked by Eve. One was blocked and the other was detected by
Bob. Both of them were detected in Bob's detection apparatus. If both of two
photons were detected in Bob's detection apparatus, Bob will discard this
qubit. Thus, we can obtain that $p_{SI}=1-p_{tagged}$.

\smallskip [15] C. A. Fuchs, N. Gisin, R. B. Griffiths, C.-S. Niu, and A.
Pere, Phys. Rev. A 56, 1163-1172 (1997).

\smallskip [16] P. Shor and J. Preskill, Phys. Rev. Lett. 85, 441-444 (2000).

[17] In decoy-state protocols, such upper bound is dependent on both $\mu $
and $\mu ^{\prime }$, $\Delta \leq \frac{\mu e^{-\mu }}{\mu ^{\prime
}e^{-\mu ^{\prime }}}$, Ref.[4,5].

[18] D. Achilles $et$ $al$., J. Mod. Opt. 51, 1499 (2004).

\section{Captions}

Caption 1. FIG. 1. Outline of Bob's detection apparatus. In order to obtain
the photon number of an arriving laser pulse, Bob first sets a group of
parallel beam splitter (BS) behind the fiber. Every BS was set at the output
of the frontal BS. Since there are two outputs of each BS, the number of BS
increases exponentially as Bob's photon number resolving power increases.
Assume the number of BS Bob used is $N$. Then the photon number resolving
power of Bob's apparatus (in A) is approximate to $O(1/N)$. The apparatus in
B will be used to obtain the states of the qubit Alice sent. A group of
parallel polarization beam splitter (PBS) was set in the output of each BS.
There is a single-photon detector (SPD) at each output of the PBS. Using
such apparatus, photon number of arriving signal can be resolved and a
precise QBER can be achieved.

Caption 2. FIG. 2. Eve's and Bob's information vs the QBER. $I^{AB}$ is
information Bob can gain with the QBER. $I^{AE}$ is the information. The
point at which privacy amplification can be implemented is that QBER=13.5\% (%
$\mu =0.1$), QBER=12.1\% ($\mu =0.2$), and QBER=10.7\% ($\mu =0.3$). Thus, a
bigger $\mu $ is allowable in our scheme. Moreover, both the signal pulses
and decoy pulses can be used to generate the raw key. Although
error-correcting can be performed as long as the QBER is lower than 11\%
[16], a small $\mu $ and a lower QBER will improve the rate of generating
the final key from the sifted key [2].

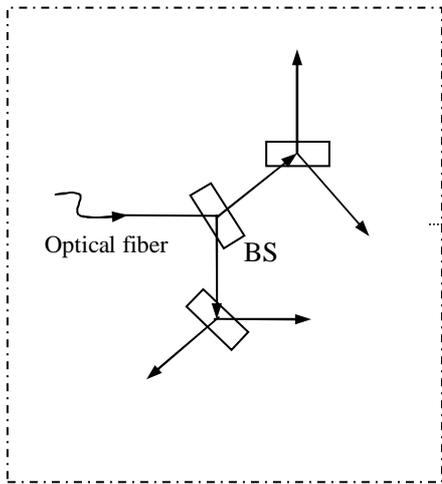
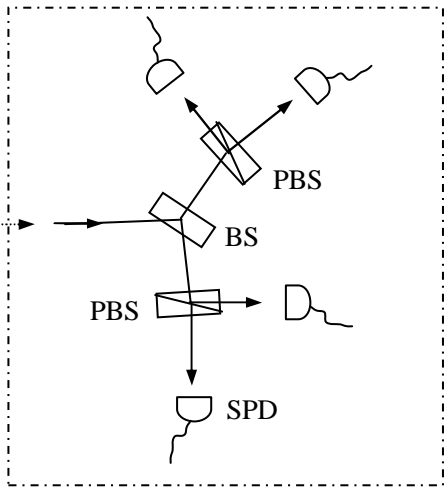
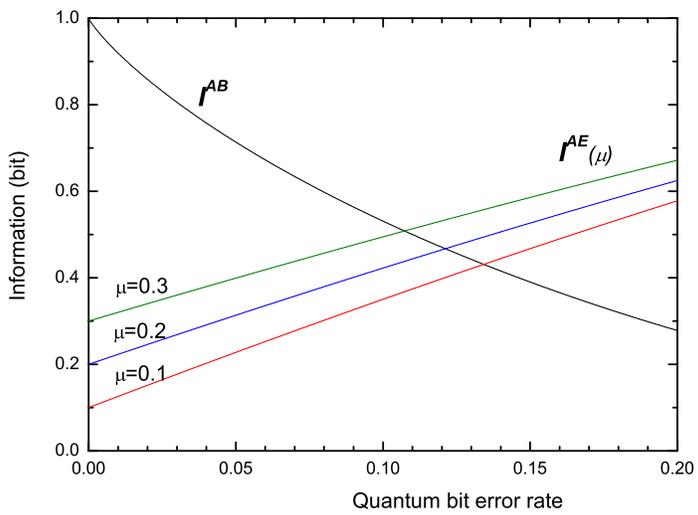
\end{document}